\documentclass[3p]{elsarticle}
\usepackage{amsmath}    
\usepackage{graphicx}   
\usepackage{color}
\usepackage[normalem]{ulem}
\usepackage{hyperref}   

\newcommand{\nuc}[2]{\textsuperscript{#1}#2}
\newcommand{\state}[2]{#1\textsuperscript{#2}}

\journal{Nuclear Physics A}

\begin{document}
    \begin{frontmatter}
        \title{Search for \nuc{21}{C} and constraints on \nuc{22}{C}}
    \author[nscl,msu]{S.~Mosby\corref{cor1}\fnref{smaddy}}
    \ead{smosby@lanl.gov}
    \author[rhodes]{N.~S.~Badger}
    \author[nscl]{T.~Baumann}
    \author[nscl]{D.~Bazin}
    \author[westmont]{M.~Bennett}
    \author[wabash]{J.~Brown}
    \author[nscl,msu]{G.~Christian\fnref{gcaddy}}
    \author[hope]{P.~A.~DeYoung}
    \author[cmu]{J.~E.~Finck}
    \author[westmont]{M.~Gardner}
    \author[iusb]{J.~D.~Hinnefeld}
    \author[rhodes]{E.~A.~Hook}
    \author[hope]{E.~M.~Lunderberg}
    \author[concordia]{B.~Luther}
    \author[rhodes]{D.~A.~Meyer}
    \author[nscl,msu]{M.~Mosby}
    \author[hope]{G.~F.~Peaslee}
    \author[westmont]{W.~F.~Rogers}
    \author[nscl,msu]{J.~K.~Smith}
    \author[nscl,msu]{J.~Snyder}
    \author[nscl,msu]{A.~Spyrou}
    \author[nscl,msu]{M.~J.~Strongman}
    \author[nscl,msu]{M.~Thoennessen}

    \cortext[cor1]{Corresponding Author.}
    \fntext[smaddy]{Present Address: LANL, Los Alamos, NM 87545, USA}
    \fntext[gcaddy]{Present Address: TRIUMF, 4004 Wesbrook Mall, Vancouver, British Columbia V6T 2A3, Canada}
    \address[nscl]{National Superconducting Cyclotron Laboratory, Michigan State University, East Lansing, Michigan 48824}
    \address[msu]{Department of Physics \& Astronomy, Michigan State University, East Lansing, Michigan 48824}
    \address[rhodes]{Department of Physics, Rhodes College, Memphis, Tennessee 38112}
    \address[westmont]{Department of Physics, Westmont College, Santa Barbara, California 93108}
    \address[wabash]{Department of Physics, Wabash College, Crawfordsville, Indiana 47933}
    \address[hope]{Department of Physics, Hope College, Holland, Michigan 49423}
    \address[cmu]{Department of Physics, Central Michigan University, Mt. Pleasant, Michigan, 48859}
    \address[iusb]{Department of Physics \& Astronomy, Indiana University at South Bend, South Bend, Indiana 46634}
    \address[concordia]{Department of Physics, Concordia College, Moorhead, Minnesota 56562}

    \begin{abstract}
        
        A search for the neutron-unbound nucleus \nuc{21}{C} was performed via the single proton removal reaction from a beam of \nuc{22}{N} at 68~MeV/u. Neutrons were detected with the Modular Neutron Array (MoNA) in coincidence with \nuc{20}{C} fragments.  No evidence for a low-lying state was found, and the reconstructed \nuc{20}{C}+n decay energy spectrum could be described with an $s$-wave line shape with a scattering length limit of $|a_s| < 2.8$~fm, consistent with shell model predictions. A comparison with a renormalized zero-range three-body model suggests that \nuc{22}C is bound by less than 70~keV.
        
    \end{abstract}

    \begin{keyword}
        Neutron Spectroscopy \sep
        Neutron Drip Line \sep
        Neutron Halo
    \end{keyword}

    \end{frontmatter}

    \section{Introduction}

    Since the discovery of the large matter radius of \nuc{11}{Li} \cite{Tanihata1985}, neutron halos have been a topic of intense study near the neutron drip line. The halo structure results from one or more valence nucleons being loosely bound which, combined with the short range of the nuclear force, allows them to have a large probability of being found at distances much greater than the normal nuclear radius \cite{Hansen1995}. For two-neutron halos the two-body subsystems are typically unbound \cite{Jensen2004} and knowledge of the basic properties of these subsystems is critical for the understanding of the halo nuclei \cite{Jensen2004,Thompson1994}.

    \begin{figure}
        \begin{center}
            \includegraphics[width=0.48\textwidth]{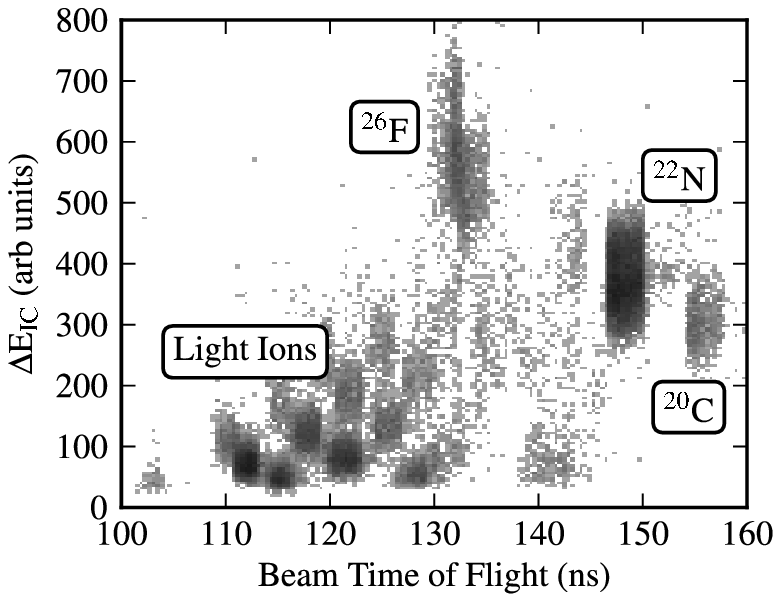}
            \includegraphics[width=0.48\textwidth]{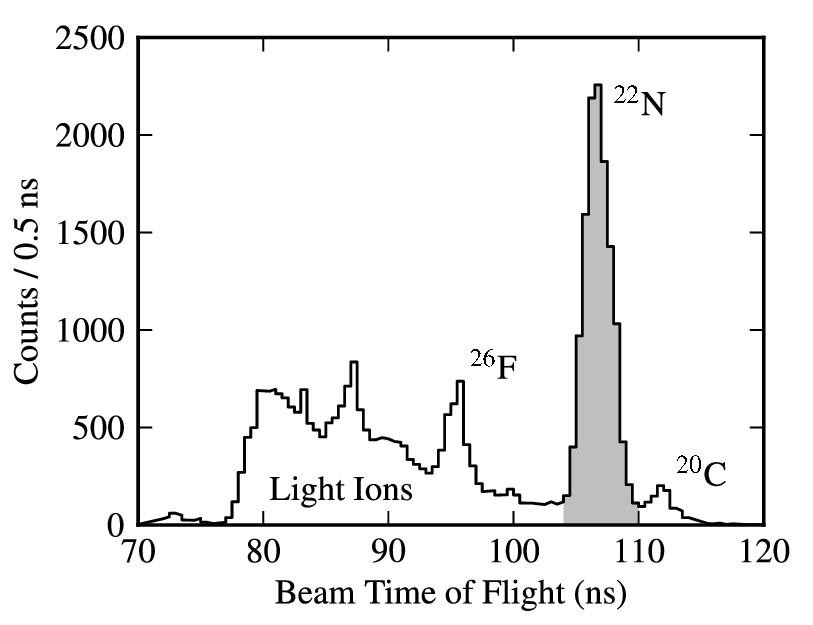}
        \end{center}
        \caption{Secondary beam composition with \nuc{22}{N} beam of interest indicated along with primary contaminants. The left panel shows the identification by energy-loss and time-of-flight. The shaded region in the right panel shows the applied gate on the time-of-flight parameter to select \nuc{22}{N}.}
        \label{fig:beam1d}
    \end{figure}

    One two-neutron halo candidate is \nuc{22}{C}, consisting of twice the number of protons and neutrons of \nuc{11}{Li}. It has attracted significant experimental  \cite{Tanaka2010,Kobayashi2012} and theoretical  \cite{Yamashita2011,Yamashita2012,Fortune2012,Ershov2012,Yuan2012} attention recently. \nuc{22}{C} was first observed to be bound in 1986 \cite{Pougheon1986}. It is a Borromean nucleus because the two-body sub-system \nuc{21}{C} had been shown to be unbound \cite{Langevin1985}, contradicting an earlier measurement which had claimed that \nuc{21}{C} was bound \cite{Stevenson1981}. It took nearly 25 years before first properties of \nuc{22}{C} were measured. A large matter radius was extracted from the measured reaction cross section, suggesting that \nuc{22}{C} exhibits a two-neutron halo \cite{Tanaka2010}. Using a simple model relating the measured radius and the two-neutron separation energy $S_{2n}$, the authors deduced a strong s-wave configuration in $^{22}$C. Subsequent momentum distribution measurements following neutron removal reactions supported this suggestion of a significant $\nu s^2_{1/2}$ valence neutron configuration in \nuc{22}{C} \cite{Kobayashi2012}. Very recently the mass excess of \nuc{22}{C} was measured to be 53.64(38)~MeV corresponding to S$_{2n} = -140(460)$~keV \cite{Gaudefroy2012}. Given the fact that \nuc{22}{C} is actually bound this value probably should be better quoted as S$_{2n} = 0^{+320}_{-0}$~keV. Theoretically, an upper limit of 220~keV for S$_{2n}$ was calculated assuming a dominating $s$-wave contribution \cite{Fortune2012}.

These observations strongly support the presence of a two-neutron halo in \nuc{22}{C}. This typically implies that the ``two-body subsystems must be either very weakly bound or low-lying resonances, or virtual states must be present, in order to support a halo state'' \cite{Jensen2004}. As mentioned earlier, \nuc{21}{C} is unbound but no spectroscopic information is presently known. The ground state of \nuc{21}{C} is expected to be a $1/2^+$ state with a large $\nu s^1_{1/2}$ single particle configuration although the possibility of a degeneracy or even level inversion with the $\nu d^1_{5/2}$ has been suggested \cite{Strongman2009}. In the present paper we present a search for a low lying resonance/virtual state in \nuc{21}{C}. We attempted to populate \nuc{21}{C} with one-proton removal reactions from a secondary beam of \nuc{22}{N} and extracted the decay-energy spectrum via invariant mass spectroscopy by measuring neutrons in coincidence with \nuc{20}{C}.

    \section{Experimental Setup}

    The experiment was performed at the National Superconducting Cyclotron Laboratory (NSCL) at Michigan State University. A 90~pnA \nuc{48}{Ca} primary beam at 140 MeV/u impinged on a 2068 mg/cm\textsuperscript{2} \nuc{9}{Be} production target, and isotopic separation of a 68 MeV/u \nuc{22}{N} secondary beam was achieved using the A1900 fragment separator \cite{Morrissey2003} with a 1057 mg/cm\textsuperscript{2} Al achromatic wedge degrader placed at the dispersive image. All data were taken with A1900 momentum slits set at 2.5\% acceptance, which resulted in a \nuc{22}{N} particle rate of 37/s with a purity of 32\%. The beam composition is shown in Figure~\ref{fig:beam1d}. The two heavy ion contaminants were \nuc{26}{F} at 6\% and \nuc{20}{C} at 2.8\%; light ions comprised the rest of the contamination. The energy loss ($\Delta$E) versus time of flight plot shown in the left panel demonstrates the clean separation of \nuc{22}{N} from the contaminates as a function of time-of-flight. The shaded area in the right panel shows the applied gate in time-of-flight to select \nuc{22}{N}.

    \begin{figure}
        \begin{center}
            \includegraphics[width=1.0\textwidth]{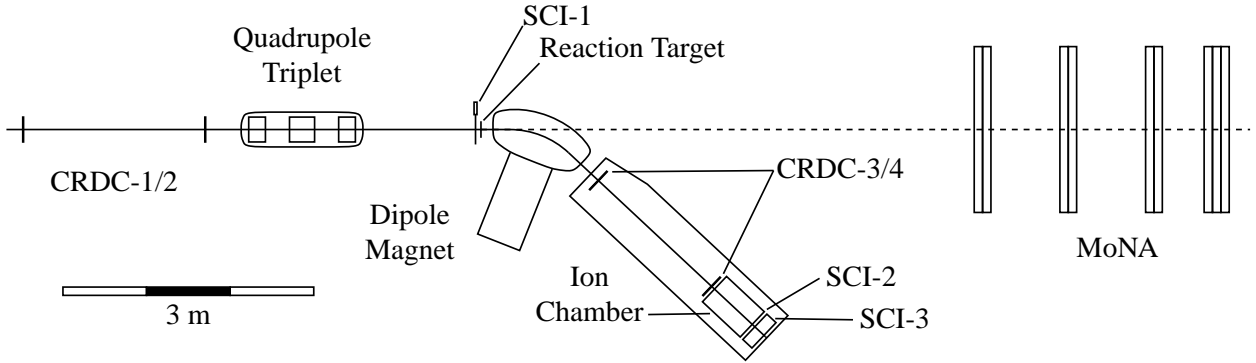}
        \end{center}
        \caption{Diagram of experimental setup.}
        \label{fig:vault}
    \end{figure}

    Figure~\ref{fig:vault} shows a diagram of the experimental setup. After exiting the A1900, the secondary beam passed through two position-sensitive cathode readout drift chambers (CRDC-1 and CRDC-2). Downstream of the CRDCs, a quadrupole triplet magnet focused the beam onto the 481~mg/cm\textsuperscript{2} \nuc{9}{Be} reaction target. A 0.254~mm plastic scintillator (SCI-1) was placed immediately upstream of the  reaction target  to determine beam, charged fragment, and neutron time of flight. The flight path length from the A1900 to the target position was 11.6~m. Neutron unbound isotopes produced in the reaction target immediately decayed into charged fragments and one or more neutrons. A large-gap superconducting dipole magnet \cite{Bird2005} bent the charged fragments away from the beam axis, and the neutrons were detected near zero degrees by the Modular Neutron Array (MoNA) \cite{Baumann2005}.

    The magnet has a maximum rigidity of 4~Tm, a vertical gap of 14~cm through which neutrons pass, and a bending angle of 43$^{\circ}$. The magnet was set to a rigidity of 3.8~Tm to center the \nuc{20}{C} fragments on the charged particle detectors. Downstream of the magnet, two 30 cm square CRDCs separated by 1.82~m provided fragment trajectory information (CRDC-3 and CRDC-4). An ionization chamber and 4.5~mm thick plastic scintillator (SCI-2) provided energy loss information for element separation, while a 150~mm thick plastic scintillator (SCI-3) provided a total residual kinetic energy measurement.

    MoNA \cite{Baumann2005} consists of 144 $10 \times 10 \times 200$~cm$^3$ plastic scintillator bars with photomultiplier tubes (PMTs) attached to each end. The modules were arranged in walls that were 16 modules tall and centered on the beam axis. Walls of 2 by 16 modules each were positioned with their front faces at 5.90 m, 6.93 m, and 7.95 m from the reaction target. A block of three walls was placed at 8.65 m. Neutron time of flight was measured from the mean time of the two PMT signals of the detector module that detected an interaction, while position across the bar was measured by the time difference between the signals.

    \section{Data Analysis}

\begin{figure}
        \begin{center}
            \includegraphics[width=0.48\textwidth]{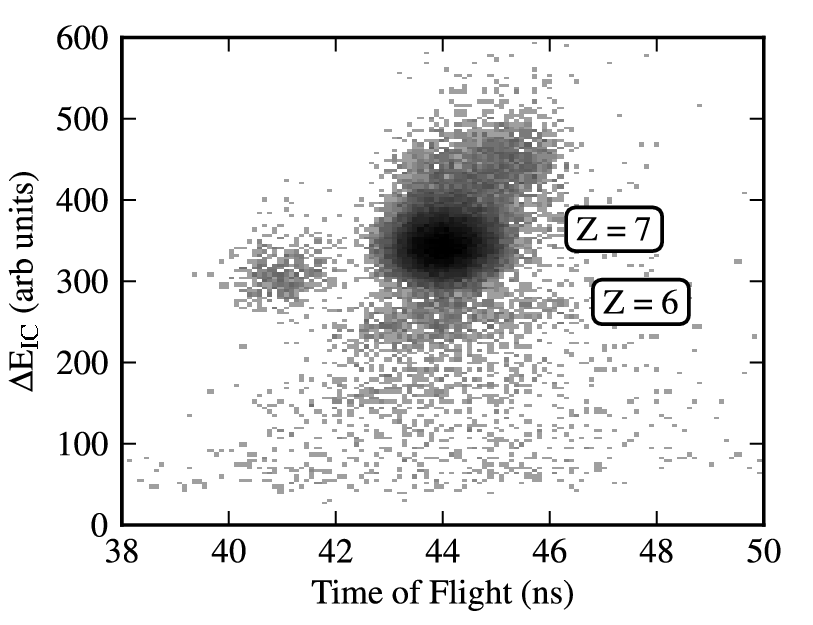}
        \end{center}
        \caption{Element separation via energy loss in the ionization chamber vs. time of flight.}
        \label{fig:eid}
    \end{figure}

\begin{figure}
        \begin{center}
            \includegraphics[width=0.48\textwidth]{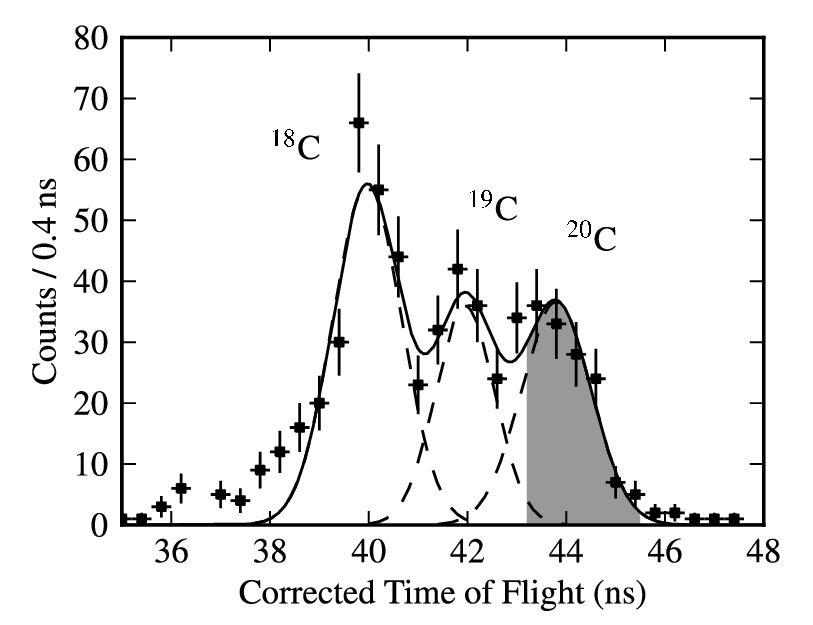}
        \end{center}
        \caption{Particle identification for neutron coincidence data showing \nuc{18-20}{C} with dashed lines demonstrating the relative contributions of the individual isotopes. The shaded region shows the selection of \nuc{20}{C}, with an estimated 2\% contamination from \nuc{19}{C} and 20\% losses from the gating procedure.}
        \label{fig:pid}
    \end{figure}

    A one-proton removal reaction was used to populate neutron unbound \nuc{21}{C} from the \nuc{22}{N} secondary beam. Charged fragments were  separated by element using energy loss information from the ionization chamber and the time of flight between SCI-1 and SCI-2  as shown in Figure~\ref{fig:eid}. Finally, isotopes of a given element were separated by correcting the time of flight between SCI-1 and SCI-2 for correlations with both dispersive and nondispersive angle and position measured by CRDC-3 and CRDC-4. This separation is shown in Figure~\ref{fig:pid}, with the grey region representing the selected \nuc{20}{C} fragments of interest. The solid line indicates the result of a fit including three convoluted Gaussian functions to estimate cross contamination between isotopes, while the dashed lines indicate each isotope's contribution to the total. With the \nuc{20}{C} selection gate shown, the contamination level is 2\% and 20\% of the \nuc{20}{C} fragments are lost.

    Reconstruction of the decay energy of \nuc{21}{C} was performed using the invariant mass method. In this method, the decay energy $E_d$ can be expressed as

    \begin{eqnarray}
        E_d = \sqrt{m_f^2 + m_n^2 + 2(E_fE_n - \vec{p_f}\cdot\vec{p}_n)} - m_f - m_n
    \end{eqnarray}

    where $E_f$ and $E_n$ are energies, $\vec{p}_f$ and $\vec{p}_n$ are momentum vectors, and $m_f$ and $m_n$ are masses for the charged \nuc{20}{C} fragment and neutron respectively.

    Energies and angles for the charged fragments at the target were constructed from the post-magnet trajectory information using a transformation matrix generated by COSY INFINITY \cite{Makino2006} based on measured magnetic field maps. The reconstruction was improved by including the measured dispersive position at the target in the transformation matrix as described in Ref. \cite{Frank2007e}. The dominant uncertainties for this process were position and angular resolution in CRDC-3 and 4, with values $\sigma_{pos} = 1.3$~mm and $\sigma_{ang} = 1$~mrad respectively as determined by data taken with a tungsten mask placed in front of the detectors.

    Neutron energies and angles were determined from the position and time of flight of the neutron interactions in MoNA. The time of flight resolution had $\sigma_{tof} = 0.3$~ns, while position resolution along a bar was characterized by shadow bar data and simulations as a sum of Laplacians \cite{Peters2007,Christian2012}

    \begin{eqnarray}
        f\frac{e^{-|x/\sigma_1|}}{2\sigma_1} + (1-f)\frac{e^{-|x/\sigma_2|}}{2\sigma_2},
    \end{eqnarray}

    where $f = 0.534$, $\sigma_1 = 16.2$~cm, and $\sigma_2 = 2.33$~cm. Discretization of the detector bars in the $y$ and $z$ directions resulted in uniform uncertainty of $\pm$5~cm for those components.

    \begin{figure}[t]
        \begin{center}
            \includegraphics[width=0.48\textwidth]{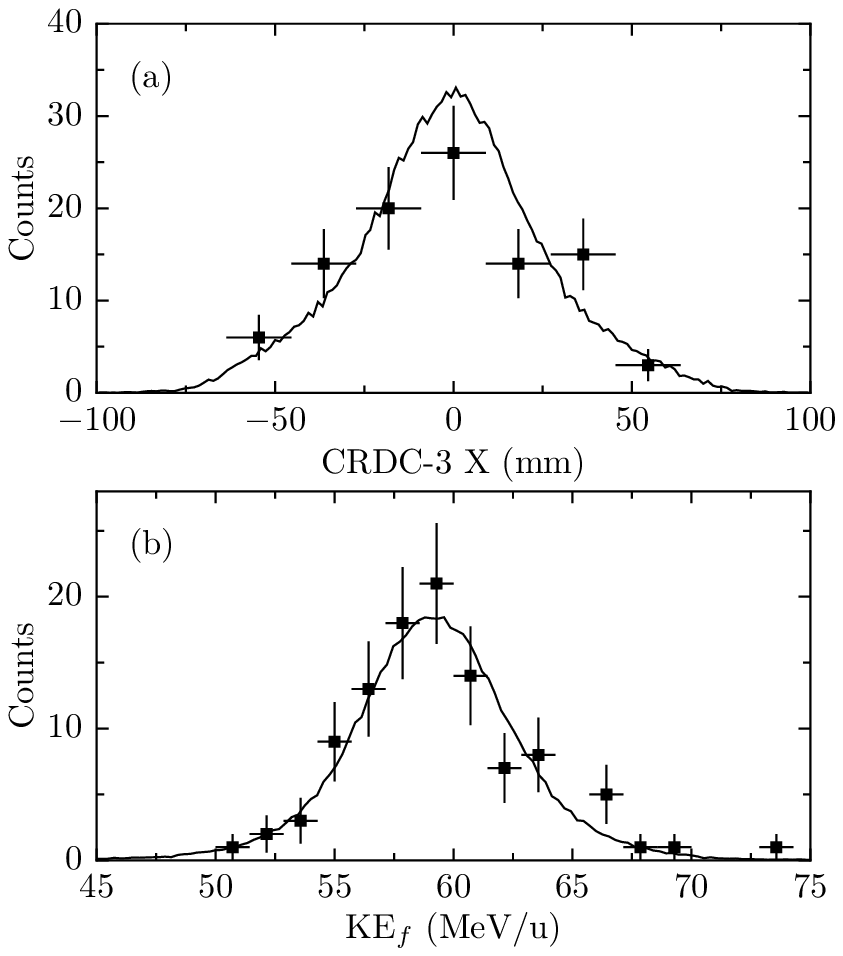}
        \end{center}
        \caption{Comparison of data (squares) with simulation (lines) for (a) \nuc{20}{C} fragment position and (b) \nuc{20}{C} fragment kinetic energy after the reaction target.}
        \label{fig:verify}
    \end{figure}

    \begin{figure}[t]
        \begin{center}
            \includegraphics[width=0.48\textwidth]{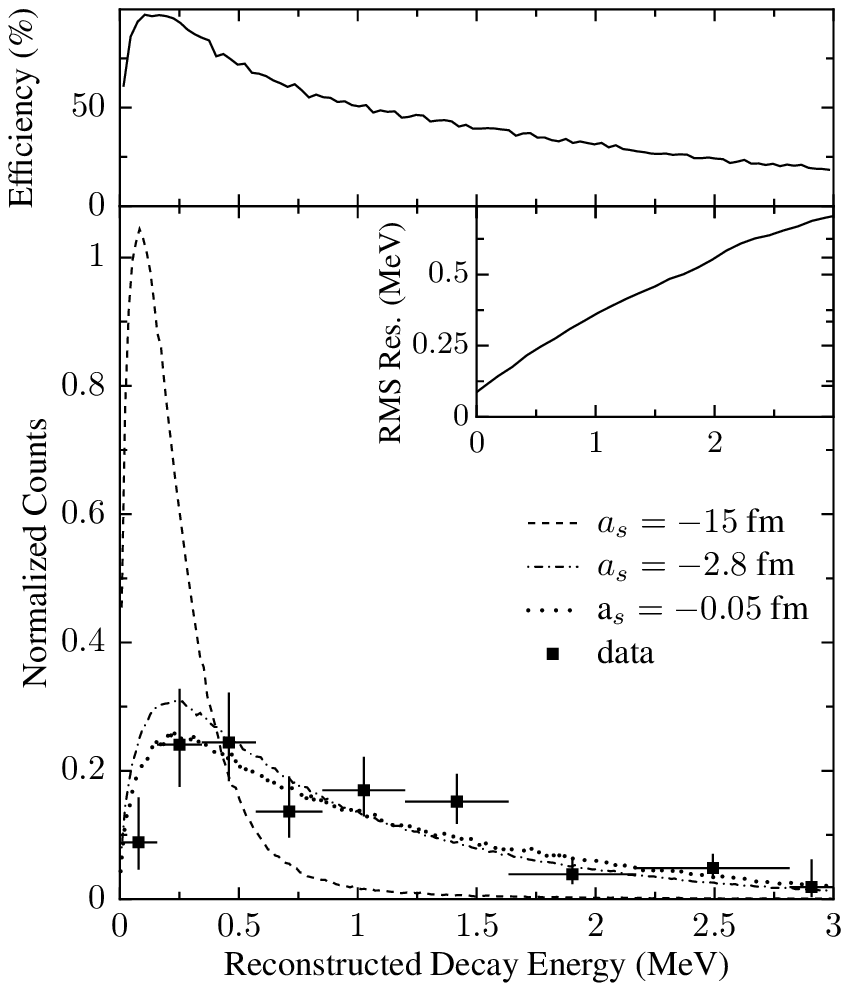}
        \end{center}
        \caption{Reconstructed decay energy spectrum (including experimental acceptance and resolution) for \nuc{21}{C} where squares represent data points. The dotted, dash-dotted, and dashed lines represent $s$-wave curves of $a_s$ = -0.05~fm, -2.8~fm, and -15~fm respectively. The inset shows the RMS decay energy resolution of the setup as a function of decay energy, and the top panel shows geometric efficiency of the setup as a function of decay energy.}
        \label{fig:edecay}
    \end{figure}

    Monte Carlo simulations of the full experimental setup, reaction process, and decay characteristics of \nuc{21}{C} were performed to account for the effects of each detector's geometrical acceptance and resolution on the reconstructed decay energy. The reaction in the target was treated by a Goldhaber model \cite{Goldhaber1974} with a friction term \cite{Tarasov2004}. These simulations were validated and all free parameters were constrained by comparison to experimental spectra except for those related to the neutron decay. Figure~\ref{fig:verify} shows the comparison between simulation and data for fragment position in CRDC-3 and reconstructed fragment kinetic energy. Simulated data sets were generated to populate grid points of the decay parameter phase space in question; these were analyzed using the same software as the experimental data and the two data sets were directly compared.

    \section{Results}

    Figure~\ref{fig:edecay} shows the measured decay energy for $^{21}$C. The data (black squares) are distributed over a broad energy range and do not exhibit any obvious resonance. A sharp resonance was not expected because the $\ell = 2$ states were not expected to be populated in the proton removal reaction from the $^{22}$N ground state, which is accepted to possess a $J^{\pi}$ of $0^-$ \cite{Sohler2008,Rodr'iguez-Tajes2011}. The calculated spectroscopic factors for the $\ell=2$ 5/2$^+$ and 3/2$^+$ excited states in $^{21}$C are 0 and 0.05, respectively. In contrast, the spectroscopic factor for populating the $\ell=0$ 1/2$^+$ state is 0.75.

    \begin{figure}
        \begin{center}
            \includegraphics[width=0.9\textwidth]{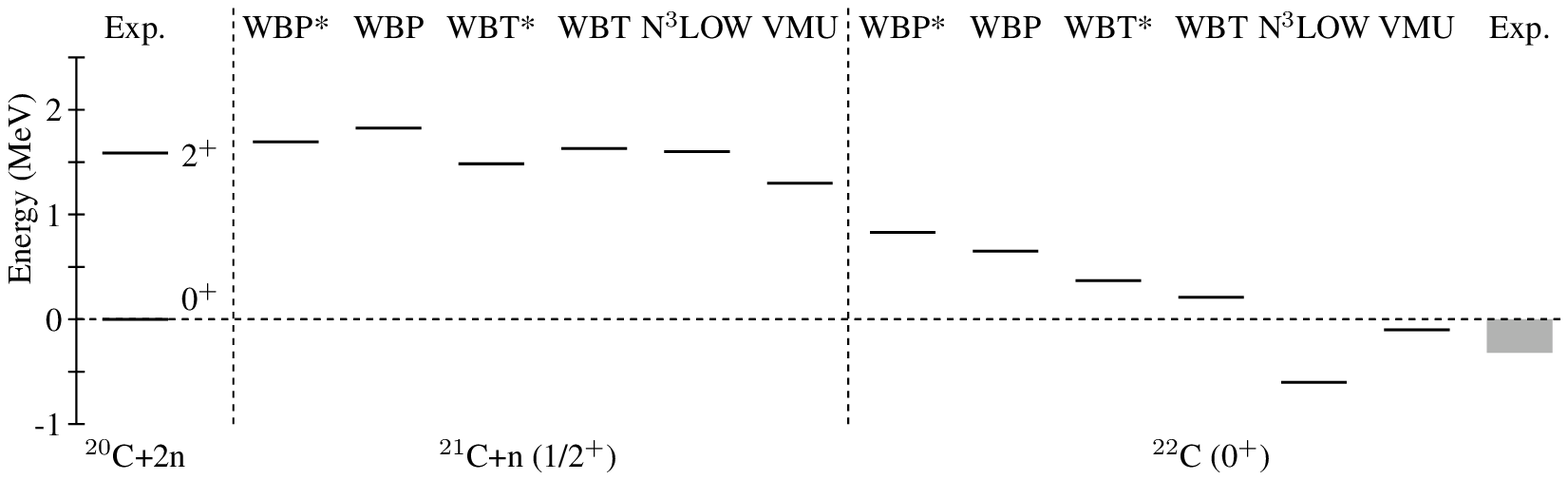}
        \end{center}
        \caption{Experimental levels of \nuc{20}{C} and theoretical $1/2^+$ and $0^+$ ground states for \nuc{21}{C} and \nuc{22}{C}, respectively. The data for \nuc{20}{C} are from \cite{Elekes2009,Stanoiu2008} and the measured two-neutron separation energy for \nuc{22}{C} from \cite{Gaudefroy2012}. The N$^3$LOW and VMU calculations are from \cite{Coraggio2010} and \cite{Yuan2012}, respectively.}
 
       \label{fig:lscheme}
    \end{figure}

    Therefore, the decay energy spectrum was fit assuming a pure $s$-wave decay, with the line shape for the decay calculated from the analytic approximation from Eq. (30) of Ref. \cite{Blanchon2007}:

    \begin{eqnarray}
        \frac{d\sigma}{d\epsilon} \sim \frac{1}{k(\gamma^2 + k^2)}\left(\frac{k\cos{\delta} + \gamma \sin{\delta}}{\sqrt{\gamma^2 + k^2}} \right)^2
    \end{eqnarray}

    where the parameterization $\delta = a_sk$ is used for the phase shift, $\gamma = \sqrt{-2m\epsilon_i}/\hbar$ is the decay length of the initial \nuc{22}{N} ground state, $\epsilon_i$ is the binding energy of \nuc{22}{N}, $k = \sqrt{2m\epsilon_f}/\hbar$ is the final momentum of the neutron in the continuum, and $\epsilon_f$ is the decay energy of the neutron.

    The scattering length $a_s$ was allowed to freely vary in the fitting process from 0 to $-$100~fm. The fitting process used a minimum $\chi^2$ technique and favored a 1$\sigma$ limit of $|a_s| < 2.8$~fm, with the best fit being $a_s = -0.05$~fm (shown in Figure~\ref{fig:edecay} as the dash-dotted and dotted lines respectively). The figure also demonstrates the sensitivity of the present set-up to any potential low energy virtual state by comparing an $s$-wave decay with $a_s = -15$~fm (dashed line) to the data and short scattering length curves. It is clear that any low lying $s$-wave state would have been apparent in the decay energy spectrum. The small magnitude of the extracted scattering length of $a_s = -0.05$~fm is essentially equivalent to no interaction at all, thus quoting a resonance energy is not meaningful \cite{Johansson2010}.

It should be mentioned that we cannot rule out the possibility of decay to the \state{2}{+} state in \nuc{20}{C} as the setup did not measure coincident $\gamma$-rays. However, it is not likely that this decay mode be observed. The spectroscopic factor for decay to the \state{0}{+} ground state of \nuc{20}{C} was calculated to be 0.38, while that for the \state{2}{+} first excited state was 1.2. Since the spectroscopic factors are relatively close in magnitude, the lack of momentum barrier for $\ell = 0$ neutron decay to the \state{0}{+} will cause it to dominate the $\ell = 2$ transition to the \state{2}{+}. Furthermore, the $\ell = 2$ transition to the \state{2}{+} would appear as a well-defined resonance in the decay energy spectrum, and no such resonant peak is observed.



    \section{Discussion}

    The results of different shell model calculations for \nuc{21}{C} and  \nuc{22}{C} are shown in Figure~\ref{fig:lscheme} together with the experimental level scheme for \nuc{20}{C} and the recently measured S$_{2n}$ for \nuc{22}{C}. We performed calculations with NuShellX@MSU \cite{Brown2011} in a truncated $s-p-sd-pf$ model space with a modified WBP \cite{Warburton1992} interaction labeled WBP*. In this interaction the neutron $sd$ two body matrix elements (TBME) were reduced to 0.75 of their original value in order to reproduce a number of observables in neutron rich carbon isotopes \cite{Stanoiu2008}. Comparison was then made to the unmodified WBP interaction as well as the WBT and WBT* interactions, where WBT* incorporates the same TBME reduction as WBP*. These interactions have been used extensively for calculations in this neutron rich region with Z$<$8 \cite{Strongman2009,Sohler2008,Stanoiu2008,Stanoiu2004a,Satou2008}. For all interactions, the \nuc{21}{C} \state{1/2}{+} $s$-state is expected to be unbound by more than 1.4~MeV, which is consistent with our measurement as no low lying states are predicted. However, as seen in the figure, all interactions also incorrectly calculate \nuc{22}{C} to be unbound.

Two recent shell model calculations with different interactions do not have the same problem. Coraggio \textit{et al.} derived the single-particle energies and the residual two-body interaction of the effective shell-model Hamiltonian from the realistic chiral NN potential N$^3$LOW \cite{Coraggio2010}. The calculations predict \nuc{21}{C} to be unbound by 1.6~MeV and \nuc{22}{C} to be bound by 601~keV which is even more bound than the recent measurement of 0$^{+320}_{-0}$~keV by Gaudefroy \textit{et al.} \cite{Gaudefroy2012}. The calculations by Yuan \textit{et al.} use a newly constructed shell-model Hamiltonian developed from a monopole-based universal interaction ($V_{MU}$) \cite{Yuan2012}. The results of the calculations are consistent with the two-neutron separation energy of \nuc{22}{C} and predict \nuc{21}{C} to be slightly less unbound ($\sim$1.3~MeV, extracted from Figure 6 of Ref. \cite{Yuan2012}) as compared to the other calculations. 

All models predict \nuc{21}{C} to be unbound by significantly more than 1~MeV. This is consistent with the non-observation of any resonance structure in the present data. However, it is somewhat unexpected as it represents the first case where the two-body subsystem of a Borromean two-neutron halo nucleus does not have a low-lying virtual state or resonance below 1 MeV. This might be another indication that the two-neutron separation energy of \nuc{22}{C} is extremely low and very close to the binding threshold.

The relationship between the two-neutron separation energy of a halo nucleus and the virtual state or resonance of the unbound two-body subsystem has been demonstrated before for \nuc{11}{Li} \cite{Thompson1994}. Higher resonance energies or smaller absolute values of the scattering length of the virtual state correspond to smaller two-neutron separation energies. Yamashita \textit{et al.} derived this relationship for the \nuc{21}{C}-\nuc{22}{C} system  within a renormalized zero-range three-body model assuming a pure $s$-wave valence neutron configuration \cite{Yamashita2011,Yamashita2012}. They establish limits on $S_{2n}$ of $\sim$120~keV and $\sim$70~keV based on the measured matter radius of \nuc{22}{C} \cite{Tanaka2010} for virtual state energies of 0~keV and 100~keV, respectively. 

For states near threshold the energy of a virtual state can be related to the scattering length with $E\simeq \hbar^2/2\mu a_s^2$, where $\mu$ is the reduced mass  \cite{Blatt1979}. This relationship places a 100~keV virtual state in \nuc{21}{C} at approximately $a_s=-15$~fm. As shown in Figure~\ref{fig:edecay} our data are not consistent with such a low-lying state, limiting the two-neutron separation energy of \nuc{22}{C} to less than 70~keV.

    \section{Conclusions}

    In summary, we have searched for the neutron-unbound nucleus \nuc{21}{C} via one-proton removal from a \nuc{22}{N} beam. The reconstructed \nuc{20}{C}+n decay energy spectrum does not contain any evidence for a low-lying state, which agrees with shell model calculations that predict this state to be neutron unbound by 1.5~MeV. From this non-observation an upper limit on the two-neutron separation energy of \nuc{22}{C} can be placed at $S_{2\text{n}} < 70$~keV based on calculations from a renormalized zero-range three-body model \cite{Yamashita2011,Yamashita2012}. This is consistent with the recently measured limit of $S_{2\text{n}} < 320$~keV \cite{Gaudefroy2012} as well as calculated limits of $S_{2\text{n}} < 220$~keV \cite{Fortune2012} and $S_{2\text{n}} < 50$~keV \cite{Ershov2012}.

In the future, the search for \nuc{21}{C} should concentrate on populating the $d_{5/2}$ state. Although the $s_{1/2}$ is still predicted to be the ground state it cannot be observed at the calculated decay energy. The $d_{5/2}$ state can be populated with a one-neutron transfer reaction with a \nuc{20}{C} beam on a deuteron target. The excitation energy is predicted to be above the first excited 2$+$ state of \nuc{20}{C} and thus $\gamma$-rays have to be detected in order to determine if any observed decay populates the ground state or the first excited state of \nuc{20}{C}.

    This work was supported by the National Science Foundation under grants PHY-05-55488, PHY-05-55439, PHY-06-51627, PHY-06-06007, PHY-08-55456, PHY-09-69173, and PHY11-02511.

    \bibliographystyle{nphys}
    \bibliography{plb_carbon_refs}
\end{document}